\documentstyle[preprint,aps]{revtex}
\tightenlines
\begin{document}
%\draft
 \begin{titlepage}
\begin{tabbing}
\hspace{11cm} \= HIP -- 1998 -- 36 / TH \\
\> \today
\end{tabbing}

\begin{centering}
\vfill
{ \Large\bf The energies and residues of the nucleon  resonances
 $N(1535)$ and $N(1650)$}

\vspace{0.5cm}

R.A~Arndt\footnotemark[1]$^, \ $\footnotemark[4]
A.M.~Green\footnotemark[2]$^, \ $\footnotemark[5]
R.L.~Workman\footnotemark[1]$^, \ $\footnotemark[6]
and S.~Wycech\footnotemark[3]$^, \ $\footnotemark[7]

{\em $\,^*$Department of Physics, Virginia Tech,
Blacksburg, VA 24061, USA  \\
$\,^{\dagger}$Department of Physics, University of Helsinki and Helsinki Institute
of Physics, P.O. Box 9, FIN--00014, Finland \\
$\,^{\ddagger}$Soltan Institute for Nuclear Studies,Warsaw, Poland}

\setcounter{footnote}{4}
\footnotetext{Email:  {\tt arndtra@said.phys.vt.edu}}
\setcounter{footnote}{5}
\footnotetext{{\tt anthony.green@helsinki.fi}}
\setcounter{footnote}{6}
\footnotetext{{\tt workman@clsaid.phys.vt.edu}}
\setcounter{footnote}{7}
\footnotetext{{\tt wycech@fuw.edu.pl}}
\renewcommand{\thefootnote}{\arabic{footnote}}
\vspace{1.5cm}

\date{\today}

\begin{abstract}
We extract pole positions for the $N(1535)$ and $N(1650)$ resonances using
two different models.  The positions are determined from fits to different
subsets of the existing $\pi N\rightarrow\pi
N$, $\pi N\rightarrow\eta N$  and $\gamma p\rightarrow\eta p$ data
and found to be 1515(10)--i85(15)MeV and 1660(10)--i65(10)MeV,
when the data is described in terms of two poles.
Sensitivity to the choice of fitted data is explored.
The corresponding $\pi \pi$ and
$\eta \eta$ residues of these poles are also extracted.
\end{abstract}


\end{centering}
\end{titlepage}
\centerline{I. INTRODUCTION}

Properties of the $N(1535)$ are difficult to extract from $\pi N\to \pi N$
and $\gamma N\to \pi N$ due to the nearby $\eta N$ threshold  \cite{new1}.
As a result, a number of recent analyses have been based on  data from
$\pi^- p\to \eta n$ and $\gamma p\to \eta p$. These studies and
coupled-channel analyses including pion production data, generally find
values for the $N(1535)$ pole position, mass, width and photo-decay
amplitudes which differ from those obtained from pion production data
alone \cite{Nefkens} - \cite{mainz}.
While these more recent studies suggest that some $N(1535)$ properties should be
revised, the modification of any single quantity is complicated due to
correlations. An example is the $\xi_p$ parameter used by Mukhopadhyay
and collaborators\cite{rpi}. This combination of the photo-decay amplitude
($A_{1/2}$), total ($\Gamma_T$) and $\eta N$ ($\Gamma_{\eta}$) widths
is relatively stable, even though values of $A_{1/2}$ and $\Gamma_T$ vary by factors
of two. Manley\cite{manley} has also noted that near-threshold
$\pi^- p\to \eta n$ data provide little sensitivity to different parameter
choices.

In Ref.~\cite{GW97} a two channel $K$-matrix model was presented for
$S$-wave $\pi N$ and $\eta N$ scattering up to a center-of-mass energy of
about 1700 MeV.
There the main motivation was  to extract the eta-nucleon scattering
length $(a)$ and effective  range $(r_0)$ and to determine their
uncertainties   allowed by the existing $\pi N\rightarrow\pi N$\cite{Arndt},
$\pi N\rightarrow\eta N$\cite{Nefkens}  and $\gamma p\rightarrow\eta p$
\cite{Krusche} data.   Below, this model will now be  used to estimate
the energies and residues of the
$S$-wave nucleon resonances $N(1535)$ and $N(1650)$ as complex poles of
the $T$-matrix.
Any  problems with the $N(1535)$ may carry over to the nearby $N(1650)$ resonance,
as the properties of these two resonances are extracted from the same
($S_{11}$) $\pi N$ partial wave and the same photoproduction multipole.

In the model of Ref.~\cite{GW97} two poles
corresponding to these resonances were included in the $K$-matrix,
and their energies were tuned along with other parameters to give a fit
to the data. However, in principle, the positions of these $T$- and $K$-matrix
poles can be quite different. Furthermore, it is the  $T$-matrix
poles that are
of physical significance -- hence their tabulation in  Ref.~\cite{PDT}.
A second reason for determining  $T$-matrix pole positions
is the greater variation of Breit-Wigner parameters
within different parameterization schemes. For each pole,
we have also extracted the corresponding residue.

The present study differs from most\cite{mainz} of those carried out
previously in that we have explored the effect of using different models and
fitting different data sets.  We have also considered,
for $\pi N$ elastic scattering, the effect of fitting
the original experimental data rather than the  amplitudes extracted from
these data. In the next section,
we compare the model used in Ref.\cite{GW97} to that used in the VPI
analyses. These two models have been utilized in our fits. In section III,
we show our results and consider the factors responsible for
differences in the extracted resonance parameters.

\centerline{II. FORMALISM}

The model of Green and Wycech is fully described in Ref.~\cite{GW97}.
Here we
repeat only the main elements, in order to facilitate comparisons
with the VPI analyses. Both models are based on a 3-channel K-matrix
formalism. In Ref.\cite{GW97}, a narrow energy range was chosen in order
to justify the neglect of partial waves beyond $l=0$. In the VPI fits,
higher partial waves were included in fits which spanned a much wider
energy range. However, these fits, while employing a multi-channel formalism,
were not constrained by  $\eta$-production data.

In the fits of Ref.\cite{GW97},
$S$-wave scattering was considered
in a system consisting of the two channels
$\pi N$ and $\eta N$ -- here denoted simply by the indices
$\pi$ and $\eta$. Then the $K$-matrix and the corresponding $T$-matrix , which are related
by $T=K+iKqT$, can be written as

\begin{equation}
\label{KT}
  K = \left( \begin{array}{ll}
 K_{\pi \pi} & K_{\eta \pi} \\
 K_{\pi \eta} & K_{\eta \eta} \end{array} \right)
 \ \ \ {\rm and} \ \
  T = \left( \begin{array}{ll}
 \frac{A_{\pi \pi}}{1-iq_{\pi}A_{\pi \pi}}&  \frac{A_{\eta
 \pi}}{1-iq_{\eta}A_{\eta \eta}}\\
  \frac{A_{\pi \eta}}{1-iq_{\eta}A_{\eta \eta}}&  \frac{A_{\eta
 \eta}}{1-iq_{\eta}A_{\eta \eta}}\end{array} \right),
\end{equation}
where $q_{\pi,\eta}$ are the center-of-mass momenta of the two mesons in the two
channels $\pi,\eta$.
The channel scattering lengths $A_{ij}$ are expressed in terms of the
$K$-matrix elements as
\begin{center}
$A_{\pi \pi}=K_{\pi \pi}+iK^2_{\pi \eta}q_{\eta}/(1-iq_{\eta}K_{\eta \eta})$,
 \ \ $ A_{\eta \pi}=K_{\eta \pi}/(1-iq_{\pi}K_{\pi \pi}) $
\end{center}
\begin{equation}
\label{2.2}
A_{\eta \eta}=K_{\eta \eta}+iK^2_{\eta \pi}q_{\pi}/(1-iq_{\pi}K_{\pi \pi}).
\end{equation}
As discussed in Ref.~\cite{GW97}, these $K$-matrices are designed to
account directly for several observed features of the experimental data
such as the presence of  two $S$-wave $\pi N$ resonances and allow
both to have a coupling to the two-pion channel. The latter
channel is not treated explicitly, but introduced by reducing a three channel
$K$-matrix  for $\pi N$,$\eta N$ and $\pi \pi N$ into the two channel form
in Eq.~\ref{KT}.
The resultant $K$-matrices in this two channel model are then as follows:
\begin{center}
$ K_{\pi\pi}\rightarrow \frac{\gamma_{\pi}(0)}{E_0-E}+
\frac{\gamma_{\pi}(1)}{E_1-E}+i\frac{K_{\pi 3}q_3K_{3 \pi}}{1-iq_3K_{33}}$ ,
 \ \ $ K_{\pi\eta}\rightarrow
 K_{\pi\eta}+\frac{\sqrt{\gamma_{\pi}(0)\gamma_{\eta}}}{E_0-E}
+i\frac{K_{\pi 3}q_3K_{3 \eta}}{1-iq_3K_{33}},$
\end{center}
\begin{equation}
\label{KKpe}
 K_{\eta\eta}\rightarrow K_{\eta\eta}+\frac{\gamma_{\eta}}{E_0-E}
+i\frac{K_{\eta 3}q_3K_{3 \eta}}{1-iq_3K_{33}},
\end{equation}
where \ \ \ \ \ \ \
$ K_{33}= \frac{\gamma_3(0)}{E_0-E}+\frac{\gamma_3(1)}{E_1-E}$ ,
 \ \ \ \ $ K_{\pi 3}= \frac{\sqrt{\gamma_{\pi}(0)\gamma_{3}(0)}}{E_0-E}
+\frac{\sqrt{\gamma_{\pi}(1)\gamma_{3}(1)}}{E_1-E},$
\begin{equation}
\label{KK33}
 K_{\eta 3}= \frac{\sqrt{\gamma_{\eta}\gamma_{3}(0)}}{E_0-E}
\end{equation}
and $q_3$ is a three-body $\pi \pi N$ phase space.
In all there were 9 parameters in the $K$-matrices and one parameter for
normalizing the photoproduction data.

In the second (VPI) approach, a Chew-Mandlestam  $K$-matrix has been
used\cite{Arndt} to couple the elastic $\pi N$ channel to two inelastic
channels, $\eta N$ and $\pi \Delta$ (in an l=2 state).
One starts with a 3x3 matrix:
\begin{equation}
 K = \left( \begin{array}{ccc}
 K_{\pi \pi} & K_{\pi \eta} & K_{\pi \Delta} \\
 K_{\pi \eta} & K_{\eta \eta} & 0 \\
 K_{\pi \Delta} & 0 & K_{\Delta \Delta} \end{array} \right) .
\end{equation}
Following the methods outlined in Ref.\cite{Arndt}, the $T$-matrix is
written in the form
\begin{equation}
T = \rho^{1/2} K ( 1 - C K )^{-1} \rho^{1/2} ,
\end{equation}
and abbreviated as $T=\rho^{1/2} {\bar T} \rho^{1/2}$. In this notation,
the elastic $T$-matrix is given by
\begin{equation}
\bar T_{\pi \pi} = { {\bar K} \over { 1 - C_{\pi \pi} \bar K} } ,
\end{equation}
where
\begin{equation}
\label{dispersion}
\bar K = K_{\pi \pi} + { {C_{\eta N} K^2_{\pi \eta} }
              \over {1 - C_{\eta N} K_{\eta \eta} } } +
               { {C_{\pi \Delta} K^2_{\pi \Delta} }
                \over {1 - C_{\pi \Delta} K_{\Delta \Delta} } } ,
\end{equation}
$C_i$ being a dispersion integral\cite{Arndt} of phase space factors over the
appropriate unitarity cut, and $\rho = {\rm Im}C$.
Inelastic channels are given by
\begin{equation}
\bar T_{\pi i} = { {\left( 1 + C_{\pi \pi} \bar T_{\pi \pi} \right) } \over
              { \left( 1 - C_i K_{ii} \right)  } } K_{\pi i} .
\end{equation}

\centerline{III. FITS TO DATA AND AMPLITUDES}

In  Ref.~\cite{GW97} the 10 parameters were determined by fitting the
$\eta$-production data of Refs.~\cite{Nefkens,Krusche} and the energy
dependent S11 $\pi N\rightarrow\pi N$ amplitudes of \cite{Arndt} over
the center-of-mass energy range  $1350\le E_{c.m} \le 1700 $ MeV.
However,
a better approach is to fit the $\pi N\rightarrow\pi N$ experimental data
directly, thus avoiding
the intermediate step of extracting
partial wave amplitudes. Since the above  $K$-matrix formalism is designed
only for  $T_{\pi \pi}(S11)$, the other partial waves are in the form
advocated in \cite{Arndt}.
The procedure is, therefore, to first fit
with this latter form all of the
$\pi N\rightarrow\pi N$ data over the full energy range (2.1 GeV)
utilized by the VPI analyses. This fit is referred to as solution VPI.
Data are then refitted, using the form of Ref.\cite{GW97},
over the energy range
$1350\le E_{c.m} \le 1700 $ MeV, along with the $\eta-$production data,
with the non-S11 amplitudes kept fixed.
In this case  only the  parameters
of the above
$K$-matrix model are  adjusted. These  fits are referred to as solutions
GW1X, where  X/100 denotes the parameter combination $q_3 \gamma_3 (0)$ related
to the  $\pi \pi N$ branching for the $N(1535)$.
The value of  X/100 was varied
from 0.00 to 0.04, thus generating solutions  GW10 to GW14.
The results of these fits are given in Table~\ref{table1}.

We also considered the effect of modifying the form used in fitting
the $\eta$ photoproduction data. As a first step, an additional
energy dependence was added.  This amounted to
replacing $A(phot)$ in Ref.~\cite{GW97} by
$A(phot)+B(phot)[E_{c.m.}-1485]/100$. However, this had little overall
effect with $B(phot)$ being an order of magnitude smaller than
$A(phot)$. A second two-parameter form
\begin{equation}
A \propto \alpha \left( 1 + iT_{\eta N} \right) +
{ \beta \over {q_{\eta} } } T_{\eta N} ,
\end{equation}
analogous to that used in
pion photoproduction  \cite{photo}, was also used. In the above,
$\alpha$ and $\beta$ were taken simply as constants. This form was
labeled  GW2X, with X retaining its earlier meaning, and was used to
generate the results presented in Table~\ref{table2}.
  The actual values for the 9 parameters are given in
Table~\ref{table0} for  GW11 and GW21 -- the solutions with the
smallest $\chi^2$. 
In Table~\ref{table01} these parameters are converted into the more 
conventional form of  Ref.~\cite{GW97}.
Here it is seen that for GW21 some of these
parameters are very different from their on-energy-shell counterparts,
whereas those for GW11 are very similar to the on-energy-shell
parameters in Table I of Ref.~\cite{GW97}. The errors quoted in this
 table from Ref.~\cite{GW97} will be used later,
when error estimates on the pole positions and residues are made.

In order to find the poles $E_P-i\Gamma_P/2$ of the $T$-matrix in
Eq.~\ref{KT}, the energy
$E$ appearing in Eqs.~\ref{KKpe} and \ref{KK33} and in the
momenta  $q_{\pi},q_{\eta}$ and $q_3$ was everywhere converted into
 $E-i\Gamma/2$.
It is a built-in
feature of the present $K$-matrix formalism that the poles
are at the same positions in all three matrix elements
$T_{\pi\pi},T_{\eta\eta},T_{\pi\eta}$. This has been checked and found to
be so within 10 keV.

The results are given in  Table~\ref{table3} and compared with the
current values in Ref.~\cite{PDT}. There it is seen that our results are
consistent with previous values -- especially  those of
Ref.~\cite{Cutkosky}. As could be expected, our error bars are smaller
due to the improvement in, and quantity of, the experimental data now being
analysed.
From Table \ \ref{table3} it is also seen that, although there is a
dependence of the pole
positions on the 2-pion branching, the differences -- for the range of
branchings considered -- are essentially covered by the statistical errors
on the positions.

In Table \ \ref{table4} a corresponding comparison has been made for the
moduli and phases  of the residues of the $T_{\pi\pi}$ poles.
This table also shows
the moduli and phases for the two $T_{\eta\eta}$ poles. Again
as a consistency check we confirm that the residues at the
$T_{\pi\eta}$ poles are simply the square root of the $T_{\pi\pi}$ and
$T_{\eta\eta}$ residues.

In addition to the above poles there is the possibility of having poles
on other Riemann sheets -- far from the physical region -- that can be
probed by systematically reversing
the signs of $q_{\pi}$ and $q_{\eta}$.  These addition poles
are quite symmetric -- a point that can be understood in the limit
where each $K_{ij}$ is a single pole $\sqrt{\gamma_i\gamma_j}/(E_0-E)$.
In this case the $T$-matrix reduces to
$T\propto [E_0-E-i\gamma_{\pi}q_{\pi}-i\gamma_{\eta}q_{\eta}]^{-1}.$

\centerline{IV. CONCLUSIONS}

In this paper we have extracted pole positions for the $N(1535)$ and $N(1650)$
resonances using two different models --  the results being given in
Table \ \ref{table3}. It is seen that the
$N(1535)$ pole positions predicted by these two models
agree within about 15 MeV, whereas some of the predictions
of the earlier models  \cite{Cutkosky}, \cite{Arndt2} and \cite{Hoehler}
are considerably different. The $N(1650)$ pole values cannot be
directly compared, as the most recent VPI fits have further poles and
zeroes. However, if one compares to the 2 S11 resonance fit of
 Ref.\cite{vpi90}, agreement with our $N(1650)$ values is much improved.
The reasons for differences can be
manifold:
a) the models used in the analysis are different,
b) different subsets of partial wave  amplitudes are fitted,
c) data versus amplitudes are fitted,
d) only  certain data sets are fitted e.g. only $\pi N\rightarrow\pi N$
or only $\pi N\rightarrow\pi N$ plus $\pi N\rightarrow\eta N$ etc.,
e) the energy ranges over which the data is fitted can differ.
We explicitly considered one such possibility in our analysis,
by including either the
S11 $\pi N$ partial-wave amplitude or the $\pi N$ data. The
N(1535) pole position
was found to be quite sensitive to this choice, shifting
about 50 MeV higher if the partial-wave amplitudes were fitted. This
sensitivity was also seen in the associated residues.

These various alternatives question the reliability of attempting to
extrapolate into the complex energy plane the $T$-matrix from a model
that only fits a limited selection of data over a limited range of energies
on the real energy axis.
In view of this, it would be desirable to have quantitative
estimates of the errors expected on these pole positions. Unfortunately,
for those fits involving directly all of the $\pi N \rightarrow \pi N$,
it is difficult to get a meaningful estimate of such errors.
However, in the less ambitious approach of Ref.~\cite{GW97}, where only
the  $\pi N \rightarrow \pi N$ S11 amplitudes -- taking into account
their error bars -- were fitted, the Minuit minimization procedure gave
error bars on the 10 parameters defining the model i.e. the $E_i$ and
$\gamma_i$ in Eqs.~\ref{KKpe},\ref{KK33}. Therefore, the errors
$\delta E_P$ and $\delta \Gamma_P/2$ on the pole positions
$E_P-i\Gamma_P/2$
could be obtained by repeating the calculation
a large number of times for a random selection of the nine
parameters defining the
$K$-matrices of eq.(\ref{KT}) -- as discussed in Ref.~\cite{GW97}.
This results in:\\
$\delta E_P(1535)\approx 10$ MeV, $\delta \Gamma_P/2(1535)\approx 10$ MeV,
$\delta E_P(1650)\approx 5$ MeV, $\delta \Gamma_P/2(1650)\approx 5$ MeV --
values that were not very dependent on the actual pole positions. Such
estimates for $\delta E_P$, $\delta \Gamma_P/2$ are
consistent with the spread of
$E_P-i\Gamma_P/2$ values from the various fits GW1X and GW2X. They are also
very close to the correlated error estimates listed in Table V.
Furthermore, it is seen that the
position of the first pole, as given in the VPI and GW models,
is consistent within these errors.
Therefore, if we were to quote a single "best" number for the pole
positions involving only two poles,
Tables~\ref{table3} and \ref{table4} suggest:\\
$E_P-i\Gamma_P/2(1535)=1510(10)-i85(15)$ and
$E_P-i\Gamma_P/2(1650)=1660(10)-i70(10)$ and  the corresponding
residues $[|r|, \theta]$ for $T_{\pi \pi}$ being
[50(10), 0(10)] and [45(10), --40(10)]. However, the residues for
$T_{\eta \eta}$ depend strongly on the fit with the components for
GW11 and GW21 differing by about a factor of two and with the VPI estimate
being somewhat closer to that of GW21.

In the above analysis the question of uniqueness arises.
In the first model, the forms
of Eqs.~\ref{KKpe} and \ref{KK33} are chosen with the physical idea in
mind that there should be two basic resonances, which are compact in
space ( as in a quark model) and so may be expected to be well represented
by a pole in the {\em K-matrix} with a constant residue. Less compact
objects would then need a form factor in place of the constant residue.
This inclusion of explicit poles in the $K$-matrix essentially guarantees
poles in the $T$-matrix in the vicinity of those  in the
$K$-matrix. In the second of our models, poles in the $K$-matrix can
arise as a dynamical effect through
coupling to high lying closed channels as in Eq.~\ref{dispersion}.
This alternative
has also been  discussed in Ref.~\cite{Kaiser}, where the $N(1535)$
is  treated as a $K\Lambda$ bound state.
This type of ambiguity has a long history and has been discussed
in most detail for the interpretation of the $\Lambda(1405)$ -- see
Refs.~\cite{Dalitz}. However, as emphasized in Ref.~\cite{Dalitz2}, the
truth is probably somewhere in between the two above possibilities,
with both mechanisms playing a role.
This seems to be supported  in Ref.~\cite{Speth}, where
the authors conclude that the $N(1535)$ is not only
generated by coupling to higher channels but "appears to require a genuine
three-quark component".
In principle, with perfect
data in all the relevant channels, the $T$-matrix should be highly
constrained, so
that only one prescription would succeed. However, in practice,
the data have error bars and only cover a limited range, so that both
approaches could give a fit to some of the data, but yield different pole
positions. As a  next step in resolving this uncertainty, all of the available
data in $\pi N\rightarrow\pi N$, $\pi N\rightarrow\eta N$  and
$\eta N\rightarrow\eta N$ (from final state interaction data in, for
example, $\gamma p\rightarrow\eta p$) should be treated simultaneously
and not simply the selections used above.

The authors  thank  Professors G.  H\"{o}hler for useful correspondence
and also one of us (S.W.) wishes to acknowledge the hospitality of the
Helsinki Institute of Physics.
The authors were saddened to hear of the untimely death of  Mijo Batini\'c,
who has played an active role in the field of $\eta$-nucleon physics.

\newpage

%\end{table}
%%%%%%%%%%%%%%%%%%%%%%%%%%%%%%%%%%%%%%%%%%%%%%%%%%%%%%%%%%%%%%%%%%
\begin{table}
\begin{center}
\caption{Dependence of the fit, using the form of
Ref.~\protect\cite{GW97}, with
variations of the branching to $\pi \pi N$. Notation is GW1X, where the
parameter combination (see text) $q_3 \gamma_3 (0)$ takes on the value
X/100. $\chi^2$ values are given for the fitted (2771) $\pi^- p$,
(452) charge-exchange (CXS), (53) $\eta$ photoproduction (Krusche),
(11) $\pi^- p\to \eta n$ total cross section (Nefkens) data.
The S11 column shows
how well this fit to data reproduces the (60) VPI S11
single-energy points. The $2\pi$ column shows the
corresponding 2-pion branching ratio as a percentage.
}
\vspace{1cm}
\begin{tabular}{cccccccc}
Soln.& Total  & $\pi^- p$ & CXS & Krusche & Nefkens & S11 & $2\pi$ \\ \hline
GW10&7719  & 6153   & 1447  & 66  & 51 & 105 & 0\\
GW11&7671  & 6169   & 1410  & 62  & 30 &  98 & 2.6\\
GW12&7718  & 6245   & 1393  & 57  & 22 & 101 & 5.2\\
GW13&7784  & 6323   & 1377  & 67  & 16 & 100 & 7.8\\
GW14&7861  & 6401   & 1374  & 73  & 13 & 104 & 10.0\\
\end{tabular}
\label{table1}
\end{center}
\end{table}
%%%%%%%%%%%%%%%%%%%%%%%%%%%%%%%%%%%%%%%%%%%%%%%%%%%%%%%%%%%%%%%%%%%%
\begin{table}
\begin{center}
\caption{Notation as in Table ~\protect\ref{table1}.
Here the two-term form (see text) has been used
to fit $\eta$ photoproduction data. The VPI solution is a fit to the
elastic $\pi N$ scattering database from
threshold to 2.1 GeV (with only forward dispersion relation constraints),
including the $\eta$-production data.
}
\vspace{1cm}
\begin{tabular}{cccccccc}
Soln.& Total  & $\pi^- p$ & CXS & Krusche & Nefkens & S11 & $2\pi$ \\ \hline
GW20&7688  & 6144   & 1442  & 51  & 50 & 126 & 0\\
GW21&7600  & 6102   & 1420  & 52  & 25 & 121 & 1.1\\
GW22&7627  & 6147   & 1410  & 53  & 17 & 122 & 2.3\\
GW23&7690  & 6227   & 1398  & 50  & 15 & 125 & 3.5\\
GW24&7774  & 6309   & 1395  & 52  & 18 & 128 & 4.8\\
VPI & 7540 & 6040  & 1397 & 53 & 49 & 85 & -- \\
\end{tabular}
\label{table2}
\end{center}
\end{table}

%%%%%%%%%%%%%%%%%%%%%%%%%%%%%%%%%%%%%%%%%%%%%%%%%%%%%%%%%%%%%%%%%%%%%

\begin{table}
\begin{center}
\caption{ The optimised parameters  defining the
$K$-matrices for GW11 and GW21 in Tables ~\protect\ref{table1}
and \protect\ref{table2}. }
\vspace{1cm}
\begin{tabular}{ccc|ccc}
&GW11 &GW21&&GW11& GW21\\ \hline
$K_{\eta \eta}$&0.1078&--0.8336&$\gamma_{\pi}$(0)&0.0640&0.1220\\
$K_{\pi \eta}$&0.0157&--0.1051&$\gamma_{\pi}$(1)&0.1071&0.0913\\
$E_0$(MeV)&1538.5&1582.5&$\gamma_{\eta}$& 0.2283&0.6027\\
$E_1$(MeV)&1681.6&1678.8&$\gamma_{3}(0)q_3$(1535)(MeV)&1.97& 1.97\\
&&&$\gamma_{3}(1)q_3$(1650)(MeV)&18.1&
16.5\\
\end{tabular}
\label{table0}
\end{center}
\end{table}

%%%%%%%%%%%%%%%%%%%%%%%%%%%%%%%%%%%%%%%%%%%%%%%%%%%%%%%%%%%%%%%%%%%%%

\begin{table}
\begin{center}
\caption{ The  parameters in Table ~\protect\ref{table0}
expressed in terms of widths and branching ratios as in Ref.
~\protect\cite{GW97}    }
\vspace{1cm}
\begin{tabular}{ccc}
&GW11 &GW21\\ \hline
$\Gamma(Total)$(MeV)&151.6&354.4\\
$\eta(br)$&0.576&0.663\\
$\pi (br)$&0.398&0.326\\
$\Gamma(Total,1)$(MeV)&150.4&133.3\\
$\pi (br,1)$&0.769&0.758\\
\end{tabular}
\label{table01}
\end{center}
\end{table}

%%%%%%%%%%%%%%%%%%%%%%%%%%%%%%%%%%%%%%%%%%%%%%%%%%%%%%%%%%%%%%%%%%%%%
\begin{table}
\begin{center}
\caption{ The Real and Imaginary parts of poles ($E_P - i\Gamma_P /2$)
in the complex energy plane compared with those quoted in
 \protect\cite{PDT}. The numbers in parentheses correspond to the 3rd
S11 pole, for which some evidence was
found in Ref.~\protect\cite{Arndt2}. }
\vspace{1cm}
\begin{tabular}{ccccc}
Reference &$E_P(1535)$(MeV)&$\Gamma_P(1535)/2$(MeV)&
$E_P(1650)$(MeV)&$\Gamma_P(1650)/2$(MeV)\\ \hline
Arndt\cite{Arndt2}&1501&62 &1673(1689)&41(96)\\
Hoehler\cite{Hoehler}&1487&--&1670&82\\
Cutkosky\cite{Cutkosky}&1510$\pm$50&130$\pm$40&1640$\pm$20&75$\pm$15\\
This paper&&&&\\
VPI & 1510$\pm 3$ & 73$\pm 3$ & 1666(1668) & 41(147)\\
VPI90\cite{vpi90} & 1499 & 55 & 1657 & 80 \\
GW10 & 1510$\pm$8 & 87$\pm$5 &      1662$\pm$3 & 70$\pm$5 \\
GW11 & 1514$\pm$9 & 90$\pm$6 &        1658$\pm$4 & 69$\pm$5 \\
GW20 & 1502$\pm$3 & 80$\pm$3 &        1667$\pm$2 & 60$\pm$4 \\
GW21 & 1509$\pm$3 & 82$\pm$4 &        1663$\pm$2 & 60$\pm$4 \\
\end{tabular}
\label{table3}
\end{center}
\end{table}
%%%%%%%%%%%%%%%%%%%%%%%%%%%%%%%%%%%%%%%%%%%%%%%%%%%%%%%%%%%%%%%%%%%%%
\begin{table}
\begin{center}
\caption{ The Moduli($|r|$)  and Phases($\theta$) of the residues of the
two poles in both $T_{\pi \pi}$ and  $T_{\eta \eta}$ compared wih
those  quoted in Ref.~\protect\cite{PDT}. Residues for the VPI 1650 MeV
resonance are not included, as the VPI fit has an added pole in this
region.
}
\vspace{1cm}
\begin{tabular}{cccccc}
$T_{ii}$&Reference &$|r|$(1535)(MeV)&$\theta$(1535)($^0$)&
$|r|$(1650)(MeV)&$\theta$(1650)($^0$)\\ \hline
$T_{\pi \pi}$&Arndt\cite{Arndt2}&31&--12 &22(72)&29(--85)\\
&Hoehler\cite{Hoehler}&--&--&39&--37\\
&Cutkosky\cite{Cutkosky}&120$\pm$40&15$\pm$45&60$\pm$10&--77$\pm$25\\
&This paper&&&&\\
$T_{\pi \pi}$&VPI&40&7&--&-- \\
& GW10 &53$\pm$10&--1$\pm$10&54$\pm$5&--43$\pm$5\\
&GW11&57&1&54&--48\\
&GW20&43$\pm$5&--10$\pm$5&42$\pm$5&--32$\pm$5\\
&GW21&45&--5&       42&--37\\
$T_{\eta\eta}$&VPI&41& --85&--&--\\
&GW10&91$\pm$20&--53$\pm$10&8$\pm$5&122$\pm$10\\
&GW11&98&--48&11&127\\
&GW20&43$\pm$5&--120$\pm$5& 6$\pm$10&14$\pm$15\\
&GW21&41&--121& 8&15\\
\end{tabular}
\label{table4}
\end{center}
\end{table}


\begin{thebibliography}{99}
\bibitem{mainz}
A similar study from the Mainz group,
within a generalized Lee model, has recently appeared:
J. Denschlag, L. Tiator, and D. Drechsel, nucl-th/9802063. This
work considers how well data can be described with and without
an $N(1535)$ resonance.
\end{thebibliography}
\end{document}